\documentclass[aps,pra,amsmath,amssymb,twocolumn,superscriptaddress,showpacs]{revtex4-1}
\usepackage{ulem}
\usepackage{amsmath,amssymb}
\usepackage{graphicx}	

\def\be{\begin{eqnarray}}   
\def\ee{\end{eqnarray}}
\def\vecb{\boldsymbol}

\begin{document}

\author{Shunsuke~A.~Sato}
\email{ssato@ccs.tsukuba.ac.jp}
\affiliation 
{Center for Computational Sciences, University of Tsukuba, Tsukuba 305-8577, Japan}
\affiliation 
{Max Planck Institute for the Structure and Dynamics of Matter, Luruper Chaussee 149, 22761 Hamburg, Germany}

\title{Frequency-resolved Microscopic Current Density Analysis of Linear and Nonlinear Optical Phenomena in Solids}

\begin{abstract}
We perform a frequency-resolved analysis of electron dynamics in solids to obtain microscopic insight into linear and nonlinear optical phenomena. For the analysis, we first compute the electron dynamics under optical electric fields and evaluate the microscopic current density as a function of time and space. Subsequently, we perform the Fourier transformation on the microscopic current density and obtain the corresponding quantity in the frequency domain. The frequency-resolved microscopic current density provides insight into the microscopic electron dynamics in real-space at the frequency of linear and nonlinear optical responses. We apply frequency-resolved microscopic current density analysis to light-induced electron dynamics in aluminum, silicon, and diamond based on the first-principles electron dynamics simulation according to the time-dependent density functional theory. Consequently, the nature of delocalized electrons in metals and bound electrons in semiconductors and insulators is suitably captured by the developed scheme.
\end{abstract}

\maketitle
\section{Introduction \label{sec:intro}}
The interaction between light and matter is one of the central topics in of physics, which has been extensively studied for a long time. When the field strength of light intensifies, the interaction of light with matter becomes nonlinear, triggering intriguing and effective phenomena \cite{PhysRevLett.7.118,boyd2020nonlinear,butcher1990elements,MOUROU2012720}. The recent advances in laser technology have facilitated experimental studies on highly nonlinear, ultrafast phenomena in solid-state systems \cite{RevModPhys.72.545,RevModPhys.81.163}, including high-order harmonic generation \cite{Ghimire2011,Ghimire2019} and optical control of electric currents \cite{Schiffrin2013,Higuchi2017}. These highly nonlinear, ultrafast phenomena have been attracting significant attention, not only from a fundamental science perspective but also for their technological implications \cite{Krausz2014,Schoetz2019}.

Despite the significance of light--matter interactions in extreme conditions, a comprehensive microscopic understanding of nonlinear optical phenomena remains elusive owing to the difficulty of extracting microscopic information from macroscopic experimental observations. Previous studies have thoroughly explored linear and relatively low-order nonlinear responses to fields based on perturbation theory \cite{PhysRevB.61.5337,RevModPhys.82.1959,PhysRevLett.115.216806,PhysRevLett.109.116601}, and a detailed explanation of the nonlinear optical phenomena based on key quantities in $k$-space (Brillouin zone), such as the Berry curvature and shift-vector, has been provided. However, a real-space insight into nonlinear optical phenomena remains to be developed.

The first-principles electron dynamics calculation in the time domain based on the time-dependent density functional theory (TDDFT) constitutes an effective approach to investigate complex nonequilibrium electron dynamics in solid-state systems \cite{PhysRevLett.52.997,PhysRevB.62.7998}. This method has been employed to gain microscopic insights into nonlinear phenomena induced by light \cite{doi:10.1063/1.4716192,PhysRevB.89.064304,PhysRevLett.118.087403,Hubener2017}. The outputs of TDDFT calculations, such as microscopic electron density and current density, contain valuable information regarding light-induced electron dynamics. For instance, the electron density in bulk titanium after laser irradiation revealed the nature of light-induced electron localization around titanium atoms \cite{Volkov2019}. Moreover, the time-averaged microscopic current density in $\alpha$-quartz under strong field irradiation showed that field-induced current flowed along the Si-O-Si bonds \cite{PhysRevLett.113.087401}. Recent developments in Fourier analysis of microscopic electron density under external fields have provided a clear, real-space understanding of optical phenomena in isolated systems, such as plasmon resonances in atoms, molecules, and large clusters \cite{Sinha-Roy2018,SinhaRoy2023}. However, such electron density analysis is not applicable for solids with delocalized electrons, such as metals, because the electron density does not accurately reflect light-induced electron dynamics. For instance, the electron density of homogeneous electron gas remains constant in both space and time under optical fields in the dipole approximation. In contrast to the electron density, the microscopic current density can directly capture the current flow dynamics of delocalized electrons, even in the homogeneous electron gas. Therefore, a theoretical scheme may offer a comprehensive description of bound and delocalized electrons in solids for the investigation of microscopic current density.

In this study, we perform the Fourier analysis on light-induced microscopic current density to gain insight into light-induced phenomena in solids. As a platform for the analysis, we employ real-time TDDFT calculation of electron dynamics in solids. To assess the effectiveness of the developed approach, we apply it to bulk aluminum, silicon, and diamond. The computed results demonstrate that in the linear response regime, the frequency-resolved microscopic current density appropriately captures delocalized electron dynamics in aluminum and bound electron dynamics in silicon. Additionally, we confirm that the frequency-resolved microscopic current density describes the nonlinear electron dynamics of bound electrons in diamond. Hence, the Fourier analysis of microscopic current density offers a powerful tool for gaining real-space insight into light-induced phenomena involving both localized and delocalized electrons in both linear and nonlinear regimes.

This paper is organized as follows. In Sec.~\ref{sec:method}, we first revisit the electron dynamics calculation using TDDFT. We subsequently perform Fourier analysis on the light-induced microscopic current density. In Sec.~\ref{sec:results}, we assess the nature of the frequency-resolved microscopic current density. For instance, we apply the developed scheme to bulk aluminum and silicon in the linear response regime. Additionally, we investigate applicability in the nonlinear regime by considering the third harmonic generation in diamond as an example. Our findings are summarized in n Sec.~\ref{sec:conclusion}.

\section{Methods \label{sec:method}}

\subsection{Real-time Electron Dynamics in Solids \label{subsec:tddft}}

Firstly, we briefly describe the first-principles electron dynamics calculations for solids based on TDDFT \cite{PhysRevLett.52.997}. The details of the method can be referred to fromprevious studies \cite{SATO2021110274}. In the TDDFT based on the Kohn--Sham scheme, the electron dynamics in solids are described by the following time-dependent Kohn--Sham equation:
\begin{align}
i \hbar \frac{\partial}{\partial t}u_{b\vecb k}(\vecb r,t) = h_{\vecb k}(t)u_{b\vecb k}(\vecb r,t),
\label{eq:tdks}
\end{align}
where $b$ is the band index, and $\vecb k$ is the Bloch wavevector. Here, $u_{b \vecb k}(\vecb r,t)$ is a periodic part of the time-dependent Bloch orbital and satisfies the periodic boundary condition, $u_{b \vecb k}(\vecb r, t)=u_{b \vecb k}(\vecb r+\vecb a_n, t)$, with the lattice vectors, $\vecb a_n$. The time-dependent Kohn--Sham Hamiltonian, $h_{\vecb k}(t)$, is defined as
\begin{align}
h_{\vecb k}(t) = \frac{1}{2m_e} \left [\vecb p + \hbar \vecb k + \frac{e}{c}\vecb A(t) \right ]^2 + \hat v_{\mathrm{ion}} + v_{\mathrm{Hxc}}(\vecb r,t),
\end{align}
where $\vecb A(t)$ is a vector potential, which is related to a spatially uniform time-dependent electric field as $\vecb E(t)=-\frac{1}{c}\dot {\vecb A}(t)$. In the present study, the electron--ion interaction $\hat v_{\mathrm{ion}}$ is described by the norm-conserving pseudopotential approximation \cite{PhysRevLett.48.1425,PhysRevB.43.1993}. Here, $v_{\mathrm{Hxc}}(\vecb r,t)$ represents the time-dependent Hartree-exchange-correlation potential, and we employ adiabatic local density approximation \cite{PhysRevB.23.5048}.

Note that one may add the exchange-correlation vector potential $\vecb{A}_{xc}$ to the vector potential $\vecb{A}(t)$ based on the time-dependent current density functional theory \cite{PhysRevLett.77.2037,PhysRevLett.79.4878,10.1063/1.481315}. In this work, we set $\vecb A_{xc}$ to zero for simplicity. 

To describe electron dynamics under specific fields $\vecb A(t)$, we solve the time-dependent Kohn--Sham equation, Eq.~(\ref{eq:tdks}), in the time domain. Note that the initial conditions for the Kohn--Sham orbitals are obtained by solving the static Kohn--Sham equation self-consistently.
\begin{align}
h_{\vecb k}(t=-\infty)u_{b\vecb k}(\vecb r,t=-\infty)=\epsilon_{b\vecb k}u_{b\vecb k}(\vecb r,t=-\infty),
\end{align}
where $\epsilon_{b\vecb k}$ is the corresponding Kohn-Sham eigenvalue.

Employing the computed time-dependent Kohn--Sham orbitals $u_{b\vecb k}(t)$, several physical observables can be evaluated. Among various observables, the macroscopic current density $\vecb J(t)$ is an important quantity for studying light--matter interactions, and it is defined as
\begin{align}
\vecb J(t) = -\frac{e}{m_e \Omega_{\mathrm{cell}}} \sum_{b} \int_{\mathrm{BZ}} d\vecb k f_{b\vecb k} 
\int_{\mathrm{cell}}d\vecb r u^*_{b\vecb k}(\vecb r,t)\vecb v_{\vecb k}(t)u_{b\vecb k}(\vecb r,t),
\label{eq:macro-current}
\end{align}
where $\Omega_{\mathrm{cell}}$ is the unit-cell volume, and $f_{b\vecb k}$ is the occupation factor that satisfies
\begin{align}
\sum_b \int_{\mathrm{BZ}} d\vecb k f_{b\vecb k} = N_{\mathrm{elec}}
\end{align}
with the number of electrons in te unit-cell, $N_{\mathrm{elec}}$. In Eq.~(\ref{eq:macro-current}), the velocity operator, $\vecb v_{\vecb k}(t)$, is defined as
\begin{align}
\vecb v_{\vecb k}(t) = \frac{1}{i\hbar}\left [\vecb r, h_{\vecb k}(t) \right ],
\end{align}
where $\vecb r$ is the position operator. The velocity operator, $\vecb v_{\vecb k}(t)$, can be decomposed into a (semi)local part and nonlocal part:
\begin{align}
\vecb v_{\vecb k}(t) = \vecb v^{\mathrm{local}}_{\vecb k}(t)+\vecb v^{\mathrm{nonlocal}}_{\vecb k}(t).
\end{align}
Here, the local and nonlocal parts are defined as follows
\begin{align}
\vecb v^{\mathrm{local}}_{\vecb k}(t) &=\frac{1}{m_e} \left [\vecb p + \hbar \vecb k + \frac{e}{c}\vecb A(t) \right ], \label{eq:velocity-local} \\
\vecb v^{\mathrm{nonlocal}}_{\vecb k}(t) &= \frac{1}{i\hbar}\left [\vecb r, \hat v_{\mathrm{ion}} \right ]. \label{eq:velocity-nonlocal}
\end{align}
Note that the nonlocal contribution originates from the nonlocal part of the pseudopotential approximation, and it vanishes in the all-electron calculations.

Various optical properties of matter can be investigated by analyzing the macroscopic current density $\vecb J(t)$ induced by an external electric field $\vecb E(t)$. For instance, linear optical properties of solids can be investigated by analyzing the macroscopic current, $\vecb J(t)$, induced by a weak perturbation \cite{PhysRevB.62.7998}. In this work, we evaluate the optical conductivity $\sigma(\omega)$ by applying the Fourier transform to the macroscopic current density $\vecb J$ that is induced by an impulsive distortion, $\vecb E(t)=\vecb e_z E_0 \delta(t)$, as
\begin{align}
\sigma(\omega)= \frac{1}{E_0}\int^{\infty}_{-\infty} dt e^{i\omega t -\gamma t} \vecb e_z \cdot \vecb J(t),
\end{align}
where $\gamma$ is a phenomenological damping parameter, and we set $\gamma$ to $0.1$~eV.

Similarly, nonlinear optical properties can be investigated from the current $\vecb J(t)$ induced by an intense laser field \cite{doi:10.1063/1.4818807,PhysRevB.94.035149,doi:10.1063/1.5068711}. Furthermore, by employing the pump--probe setup, one can investigate transient optical properties of solids in the presence of laser fields in the time domain \cite{PhysRevB.89.064304}. Hence, the macroscopic current density $\vecb J(t)$ plays a crucial role in the investigation of the optical properties of solids.

\subsection{Frequency-resolved Microsocpic Current Density \label{subsec:transition-current-density}}

We perform microscopic current density analysis in the frequency domain to gain microscopic insight into light-induced phenomena. Here, we consider the following microscopic current density
\begin{align}
\vecb j(\vecb r, t) = -\frac{e}{m_e} \sum_{b} \int_{\mathrm{BZ}} d\vecb k f_{b\vecb k} 
 u^*_{b\vecb k}(\vecb r,t)\vecb v^{\mathrm{local}}_{\vecb k}(t)u_{b\vecb k}(\vecb r,t).
\label{eq:micro-current}
\end{align}
Note that the local current density cannot be straightforwardly defined for the nonlocal part of the velocity operator, $\vecb v^{\mathrm{nonlocal}}_{\vecb k}(t)$. Furthermore, the contribution from the nonlocal operator to the macroscopic current density, $\vecb J(t)$, vanishes in the all-electron calculations. Hence, in this study, we consider the contribution only from the local part of the velocity operator, $\vecb v^{\mathrm{local}}_{\vecb k}(t)$, to define the microscopic current density, $\vecb j(\vecb r, t)$, in Eq.~(\ref{eq:micro-current}). When the contribution of the nonlocal operator, $\vecb v^{\mathrm{nonlocal}}_{\vecb k}(t)$, is negligible, the spatial average of the microscopic current density, $\vecb j(\vecb r,t)$, in the unit-cell reproduces the macroscopic current density.
\begin{align}
\vecb J(t) \approx \frac{1}{\Omega_{\mathrm{cell}}}\int_{\mathrm{cell}}d\vecb r
\vecb j(\vecb r,t).
\end{align}
Hence, the microscopic current density, $\vecb j(\vecb r,t)$, is expected to provide microscopic insight into light-induced phenomena in solids.

Having established the connection between the optical properties of solids and microscopic current density via the macroscopic current density, we further analyze the microscopic current density in the frequency domain. For this analysis, we introduce \textit{frequency-resolved microscopic current dneisty}, $\tilde {\vecb j}(\vecb r,\omega)$:
\begin{align}
\tilde {\vecb j}(\vecb r,\omega) = \int^{\infty}_{-\infty}dt W(t)e^{i\omega t} \vecb j(\vecb r,t),
\label{eq:freq-current-density}
\end{align}
where $W(t)$ is a window function for the reduction of the numerical error due to the finite simulation time. For practical calculations in this study, we employ the following window function:
\begin{align}
W(t) =\cos^2 \left [\pi \frac{t}{T_{\mathrm{pulse}}} \right ]
\end{align}
in the duration $-T_{\mathrm{pulse}}/2<t<T_{\mathrm{pulse}}/2$ and zero outside. Here, $T_{\mathrm{pulse}}$ is the full duration of applied laser pulses, which will be introduced later along with the pulse shape of the applied fields.

By analyzing the frequency-resolved microscopic current density, $\tilde {\vecb j}(\vecb r,\omega)$, in Eq.~(\ref{eq:freq-current-density}), we can gain atomic-scale insight into linear and nonlinear optical phenomena in solids at a specific frequency of optical responses. For instance, we can study whether localized electrons on specific bonds or atoms are responsible for a certain optical response or whether delocalized electrons play a central role by visualizing the frequency-resolved microscopic current density, $\tilde {\vecb j}(\vecb r,\omega)$.

The discrepancy between the Fourier transform of current density, $\vecb j(\vecb r,t)$, and that of electron density, $\rho(\vecb r,t)$, from the perspective of optical responses of solids is noteworthy. If the nonlocal part of the velocity operator is negligible, the microscopic current density and the electron density are related via the continuity equation.
\begin{align}
\frac{\partial \rho(\vecb r,t)}{\partial t} + \vecb{\nabla}\cdot \vecb j(\vecb r,t)=0.
\label{eq:continuity-eq}
\end{align}
Although certain information pertaining to the microscopic current density, $\vecb j(\vecb r,t)$, can be reproduced based on the electron density, $\rho(\vecb r,t)$, via Eq.~(\ref{eq:continuity-eq}), the divergence-free part of $\vecb j(\vecb r,t)$ cannot be reproduced, since such component does not contribute to the continuity equation. Therefore, when the divergence-free component of $\vecb j(\vecb r,t)$ plays a crucial role in optical response, the electron density may not contain sufficient information to describe such response. For instance, let us consider an ideal noninteracting Fermi gas as an approximated electronic system. The ground state electron density of such a system is constant in real-space and constitutes a sphere in momentum space, the so-called Fermi sphere. We consider applying a homogeneous electric field to such a system. Owing to the field application, the system exhibits homogeneous current density in realsspace due to the shift of the Fermi sphere in momentum space, while the electron density remains constant in both space and time. Hence, the electron density does not reflect the light-induced electron dynamics of ideal Fermi gases. This example indicates that electron density analysis is not sufficient for analyzing the optical responses of delocalized electrons, such as free electrons in metals, although density analysis is useful for isolated systems. In contrast, the microscopic current density can be used for both localized and delocalized electrons because it is directly linked to the macroscopic current density $\vecb J(t)$ and optical responses via the Maxwell equation. Hence, the microscopic current density affords a more comprehensive description of the optical responses of matter.

\section{Results \label{sec:results}}

To assess the developed approach in Sec.~\ref{subsec:transition-current-density}, we perform Fourier analysis of the light-induced microscopic current density corresponding to linear and nonlinear optical phenomena in solids. The aim of the analysis is to connect the macroscopic optical properties of solids, which have been heavily computed with various methods \cite{RevModPhys.74.601,10.1063/1.481315,PhysRevB.78.121201}, and the microscopic information provided by the current density analysis. As practical examples, we investigate linear optical responses of bulk aluminum and silicon. Furthermore, we analyze the third harmonic generation from diamond as an example of nonlinear phenomena. For practical calculations in this study, we employ a first-principles electron dynamics simulation code, SALMON~\cite{NODA2019356}, which employs real-space grid representation for the description of Kohn--Sham orbitals, $u_{\vecb k}(\vecb r,t)$.

\subsection{Linear Responses of a Metal: Aluminum \label{subsec:aluminum}}

First, we examine the microscopic current density analysis based on linear responses of a simple metal, aluminum. For the description of bulk aluminum, we employ a cubic unit cell that contains four aluminum atoms. The aluminum unit cell is discretized into $16^3$ real-space grid points, and the corresponding first Brillouin is also discretized into $24^3$ $k$-points. Aluminum atoms are described by a norm-conserving pseudopotential method, treating $3s$ and $3p$ electrons as valence electrons.

\begin{figure}[htbp]
  \centering
  \includegraphics[width=0.95\columnwidth]{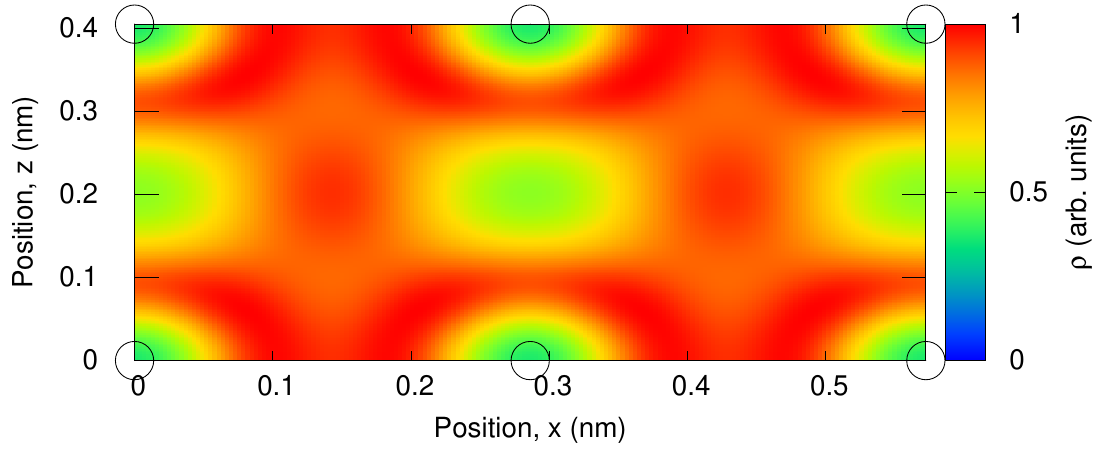}
\caption{\label{fig:rho_gs_Al}
Ground state electron density $\rho(\vecb r)$ of aluminum in the $\left  ( 110 \right )$ plane. The black circles indicate the positions of aluminum atoms. Note that contributions only from valence electrons are included, but the core-electron contribution is excluded due to the pseudopotential approximation.
}
\end{figure}

To gain insight into electron dynamics in simple metals, we revisit the electron density in aluminum. Figure~\ref{fig:rho_gs_Al} shows the computed electron density, $\rho(\vecb r)$, in the ground state of aluminum. Electrons are delocalized in aluminum. These delocalized electrons correspond to conduction electrons, and they are expected to play a central role in the optical responses of aluminum. 

\begin{figure}[htbp]
  \centering
  \includegraphics[width=0.95\columnwidth]{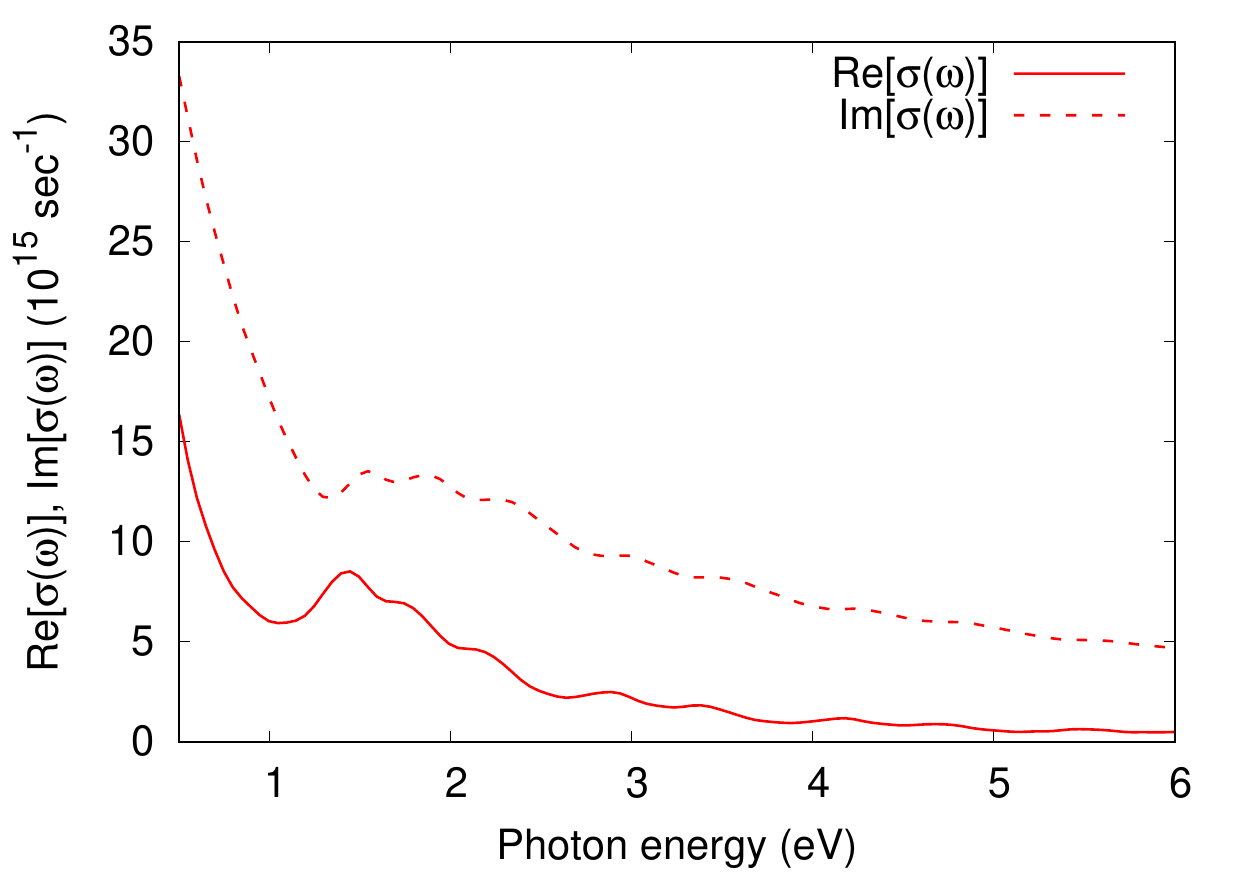}
\caption{\label{fig:sigma_Al}
Optical conductivity, $\sigma(\omega)$, of aluminum computed by TDDFT.
}
\end{figure}

Figure~\ref{fig:sigma_Al} shows the optical conductivity $\sigma(\omega)$ of bulk aluminum computed via TDDFT calculations. Note that the conductivity and dielectric function, $\epsilon(\omega)$, are connected via $\epsilon(\omega)=1+\frac{4\pi i}{\omega}\sigma(\omega)$, and the photoabsorption is closely related to the real part of the conductivity, $\mathrm{Re}\left [\sigma(\omega) \right ]$, and the imaginary part of the dielectric function, $\mathrm{Im} \left [\epsilon(\omega) \right ]$. As shown in Fig.~\ref{fig:sigma_Al}, both real and imaginary parts of the conductivity show positive values in the investigated photon energy range. This is a typical metallic response with respect to optical fields below the plasma frequency, and the Drude model can describe it suitably.

To investigate the metallic response of aluminum, we compute electron dynamics under a laser pulse described by the following vector potential,
\begin{align}
\vecb A(t) = -\vecb e_z cE_0 \cos^2 \left [\pi \frac{t}{T_{\mathrm{pulse}}} \right ] \sin \left (\omega_{L}t \right )
\label{eq:vec-pot-al}
\end{align}
in the duration $-T_{\mathrm{pulse}}/2<t<T_{\mathrm{pulse}}/2$ and zero outside. Here, $\vecb e_z$ is the unit vector along $z$-axis, which is $(001)$-direction of the unit cell, $E_0$ is the peak field strength, $T_{\mathrm{pulse}}$ is the full pulse duration, and $\omega_L$ is the mean frequency of the laser field. For the calculations of electron dynamics in aluminum, we set $T_{\mathrm{pulse}}$ to $20$~fs, $\omega_L$ to $1.55$~ev/$\hbar$, and $E_0$ to $2.75\times 10^{8}$~V/m. Employing the computed time-dependent current density, $\vecb j(\vecb r,t)$, we further evaluate the frequency-resolved microscopic current density using Eq.~(\ref{eq:freq-current-density}) based on the electron dynamics calculation.

\begin{figure}[htbp]
  \centering
  \includegraphics[width=0.95\columnwidth]{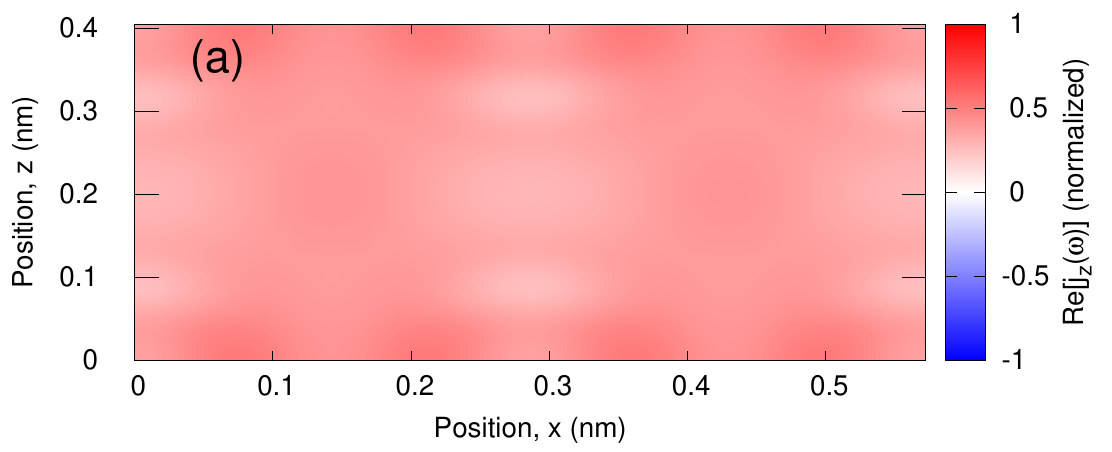}
  \includegraphics[width=0.95\columnwidth]{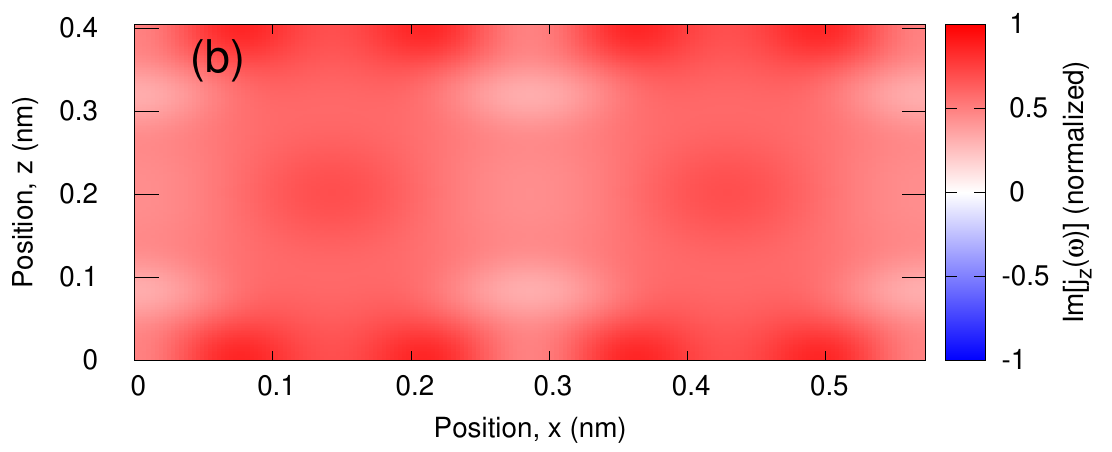}
\caption{\label{fig:zjz_re_w_Al}
Frequency-resolved microscopic current density in aluminum: $\tilde{j}_z(\vecb r,\omega)$. The real part is shown in (a), while the imaginary part is shown in (b). The corresponding electron dynamics is computed under the external field, Eq.~(\ref{eq:vec-pot-al}), and $\tilde{\vecb j}(\vecb r,\omega)$ is evaluated at the mean frequency of the driving field, $\omega=\omega_L$. Here, $\tilde{j}_z(\vecb r,\omega_L)$ is normalized by the maximum value of $\left |\tilde{j}_z(\vecb r,\omega_L) \right |$ in the entire space.
}
\end{figure}

Figure~\ref{fig:zjz_re_w_Al} shows the $z$-component of the frequency-resolved microscopic current density, $\tilde{\vecb j}(\vecb r,\omega=\omega_L)$, at the driving laser frequency $\omega_{L}$. Since the carrier wave of the electric field $\vecb E(t)$ is derived from the vector potential, Eq~(\ref{eq:vec-pot-al}) and it is proportional to $\cos(\omega_L t)$, the real part of $\tilde{\vecb j}(\vecb r,\omega_L)$ corresponds to the real part of the conductivity in Fig.~\ref{fig:sigma_Al}, while the imaginary part of $\tilde{\vecb j}(\vecb r,\omega_L)$ corresponds to the imaginary part of the conductivity.

As observed in Fig.~(\ref{fig:zjz_re_w_Al}), both real and imaginary parts of $\tilde{j}_z(\vecb r,\omega_L)$ in aluminum exhibit quasi-homogeneous positive contributions to the whole space. This indicates that the optical responses of simple metals are dominated by delocalized free electrons in the conduction bands, and the frequency-resolved microscopic current density suitably captures this metallic nature of the optical responses with free carriers.

\subsection{Linear Responses of a Semiconductor: Silicon \label{subsec:silicon}}

We subsequently perform Fourier analysis on the microscopic current density with linear responses of a typical semiconductor, i.e., silicon. For the description of silicon, we employ a cubic unit cell that contains eight silicon atoms. The silicon unit cell is discretized into $20^3$ real-space grid points, and the corresponding first Brillouin zone is also discretized into $24^3$ $k$-points. Silicon atoms are described via a norm-conserving pseudopotential method, treating $3s$ and $3p$ electrons as valence electrons.

\begin{figure}[htbp]
  \centering
  \includegraphics[width=0.95\columnwidth]{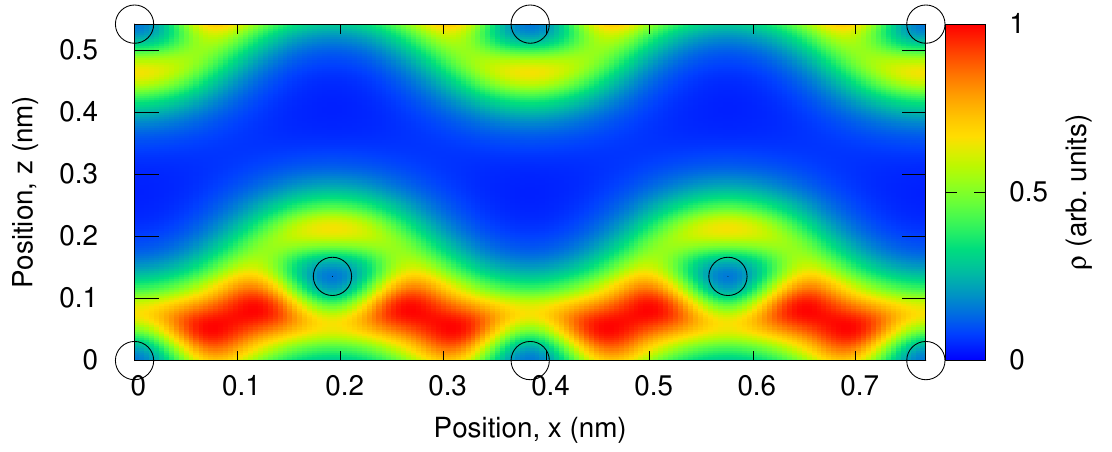}
\caption{\label{fig:rho_gs_Si}
Ground state electron density $\rho(\vecb r)$ of silicon on the $\left  ( 110 \right )$ plane. The black circles indicate the positions of silicon atoms. Note that contributions only from valence electrons are included, but the core-electron contribution is excluded due to the pseudopotential approximation.
}
\end{figure}

To obtain microscopic insight into the electronic properties of silicon, we revisit the electron density of silicon. Figure~\ref{fig:rho_gs_Si} shows the computed electron density $\rho(\vecb r)$ of silicon in the ground state. The electron density is concentrated between the two silicon atoms, forming covalent bonds. In contrast to delocalized electrons in aluminum in Fig.~\ref{fig:rho_gs_Al}, electrons are highly localized in silicon. Hence, dielectric responses of silicon are expected to be dominated by the dynamics of localized electrons at covalent bonds.

\begin{figure}[htbp]
  \centering
  \includegraphics[width=0.95\columnwidth]{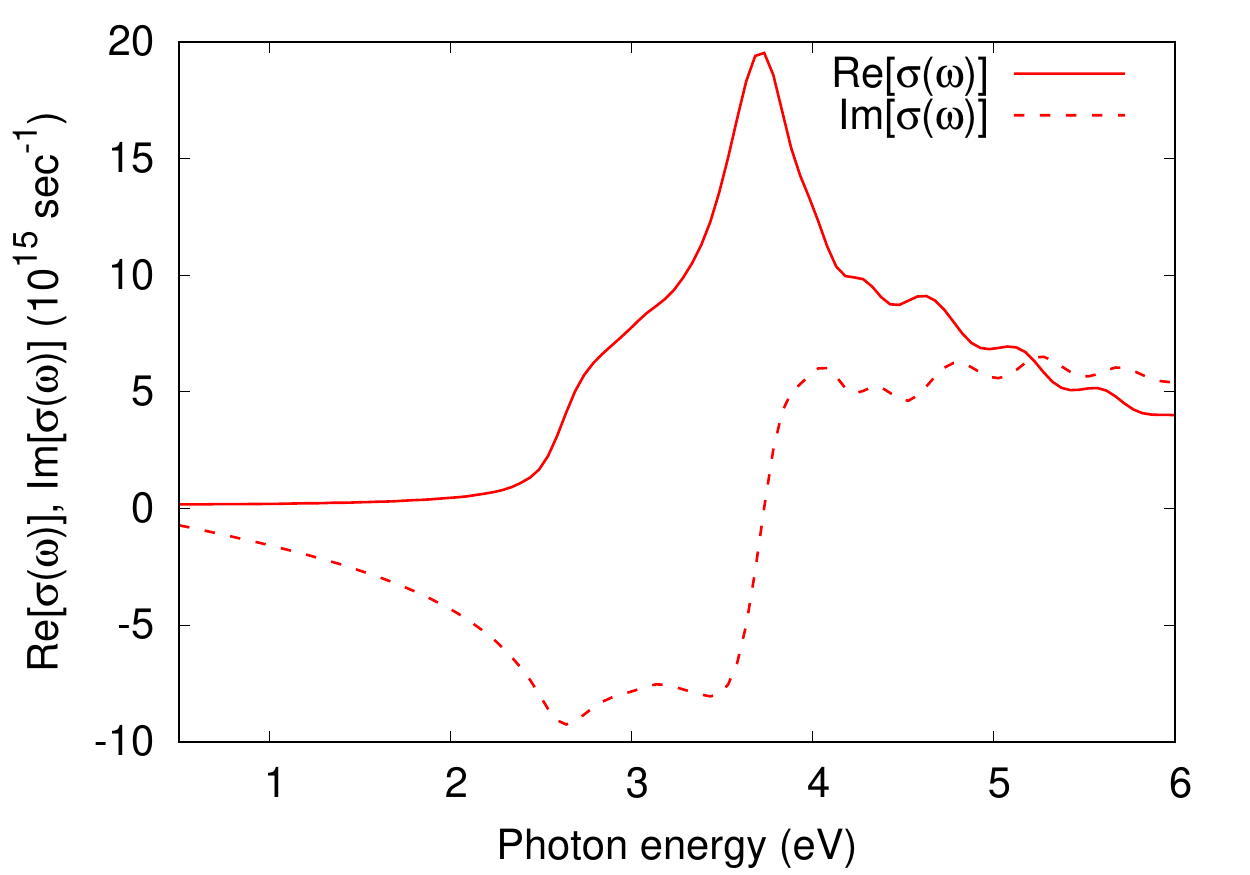}
\caption{\label{fig:sigma_Si}
Optical conductivity, $\sigma(\omega)$, of silicon computed by TDDFT.
}
\end{figure}

Here, we elucidate the dielectric responses of silicon. For this purpose, we compute the optical conductivity of silicon. Figure~\ref{fig:sigma_Si} shows the real and imaginary parts of the optical conductivity, $\sigma(\omega)$, of silicon computed through TDDFT. The computed optical gap of silicon is approximately $2.4$~eV, and the real part of the conductivity almost vanishes below the optical gap, reflecting the absence of photoabsorption below the gap. In contrast to the aluminum case in Fig.~\ref{fig:sigma_Al}, the imaginary part of the conductivity of silicon becomes negative below the optical gap. The positive value of the imaginary part of the conductivity of aluminum in the low-frequency region, i.e., at $1.55$~eV, can be explained in terms of the response of free carriers using the Drude model, whereas the negative value of the conductivity of silicon in the low-frequency regime can be explained in terms of the response of bound electrons using the Lorentz model, which is a classical model based on harmonic oscillators. This also indicates that the optical responses of silicon in the low photon energy region below the optical gap are dominated by the bound electrons at the covalent bonds.

Note that the computed optical gap of silicon, $2.4$ eV, underestimates the experimental value of $3.4$ eV \cite{PhysRevB.27.985}. This underestimation is caused by the local density approximation used in the present study. Nevertheless, this underestimation of the gap does not affect the conclusions of this research because the scope of this study is to establish a scheme for analyzing the optical responses of solids and demonstrate its feasibility. Note that the proposed microscopic current density analysis can be combined with any theoretical methods as long as the microscopic current density $\vecb j(\vecb r,t)$ is provided. Hence, the frequency-resolved current density analysis can be easily extended to TDDFT calculations with more advanced exchange-correlation functionals, which correct the gap underestimation problems. Furthermore, one may apply the microscopic current density analysis even to the time-dependent current density functional theory and the exact many-body Schr\"odinger equation.

\begin{figure}[htbp]
  \centering
  \includegraphics[width=0.95\columnwidth]{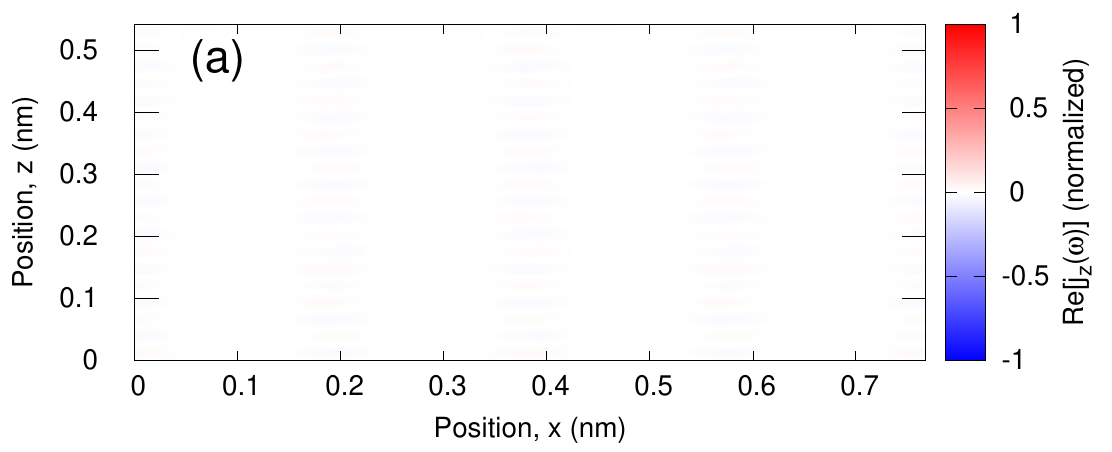}
  \includegraphics[width=0.95\columnwidth]{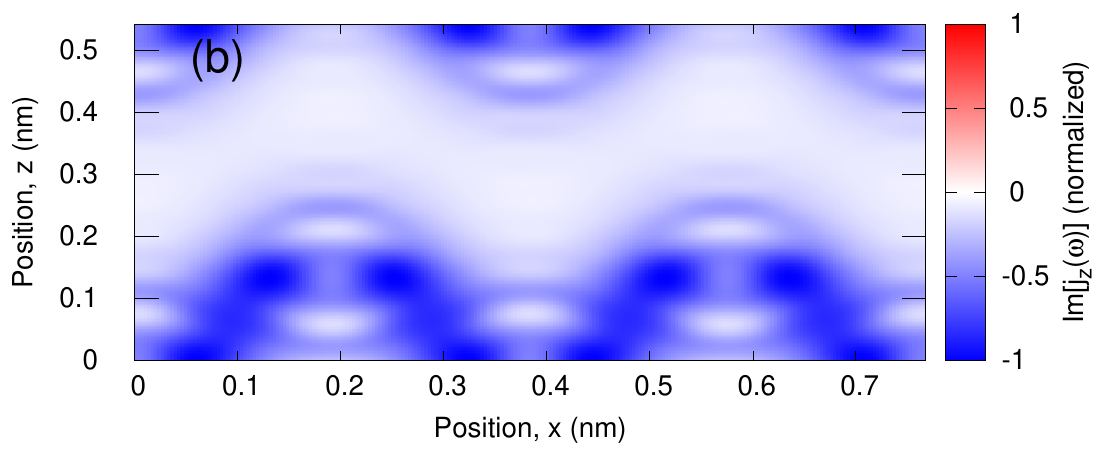}
\caption{\label{fig:zjz_re_w_Si}
Frequency-resolved microscopic current density in silicon: $\tilde{j}_z(\vecb r,\omega)$. The real part is shown in (a), while the imaginary part is shown in (b). The corresponding electron dynamics is computed under the same external field shown in Fig.~\ref{fig:zjz_re_w_Al}, and $\tilde{\vecb j}(\vecb r,\omega)$ is evaluated at the mean frequency of the driving field, $\omega=\omega_L$. Here, $\tilde{j}_z(\vecb r,\omega_L)$ is normalized by the maximum value of $\left |\tilde{j}_z(\vecb r,\omega_L) \right |$ in the entire space.
}
\end{figure}

To develop microscopic insight into the optical response of silicon, we compute the electron dynamics under the field described with Eq.~(\ref{eq:vec-pot-al}). Here, we use the same parameters as the analysis of aluminum in Sec.~\ref{subsec:aluminum}. We subsequently evaluate the frequency-resolved microscopic electron density at the frequency of the laser field as $\tilde{\vecb j}(\vecb r,\omega=\omega_L)$ with $\omega_L=1.55$~eV/$\hbar$. Figure~\ref{fig:zjz_re_w_Si} shows the $z$-component of $\tilde{\vecb j}(\vecb r,\omega_L)$.  Consistent with the optical conductivity at $\omega_L=1.55$~eV/$\hbar$, the real part of the frequency-resolved microscopic current density is almost zero in Fig~\ref{fig:zjz_re_w_Si}~(a), while the imaginary part shows negative values. Furthermore, by comparing Fig.~\ref{fig:zjz_re_w_Si}~(b) with Fig.~\ref{fig:rho_gs_Si}, the microscopic current density of silicon is found to be concentrated in the covalent bond region. Hence, the frequency-resolved microscopic current density analysis indicates that the optical responses of silicon below the optical gap are dominated by the responses of the bound electrons localized at the covalent bonds.

\subsection{Non-linear Responses of an Insulator: Diamond \label{subsec:diamond}}

Having demonstrated that the frequency-resolved microscopic current density analysis can capture both metallic responses of delocalized free carriers and dielectric responses of localized electrons at covalent bonds, we subsequently apply Fourier analysis to nonlinear optical responses. As an example for the demonstration, we consider the third harmonic generation from diamond. For the description of diamond, we employ a cubic unit cell that contains eight carbon atoms. The diamond unit cell is discretized into $20^3$ real-space grid points, and the corresponding first Brillouin zone is also discretized into $24^3$ $k$-points. Carbon atoms are described by a norm-conserving pseudopotential method, treating $2s$ and $2p$ electrons as valence electrons.

\begin{figure}[htbp]
  \centering
  \includegraphics[width=0.95\columnwidth]{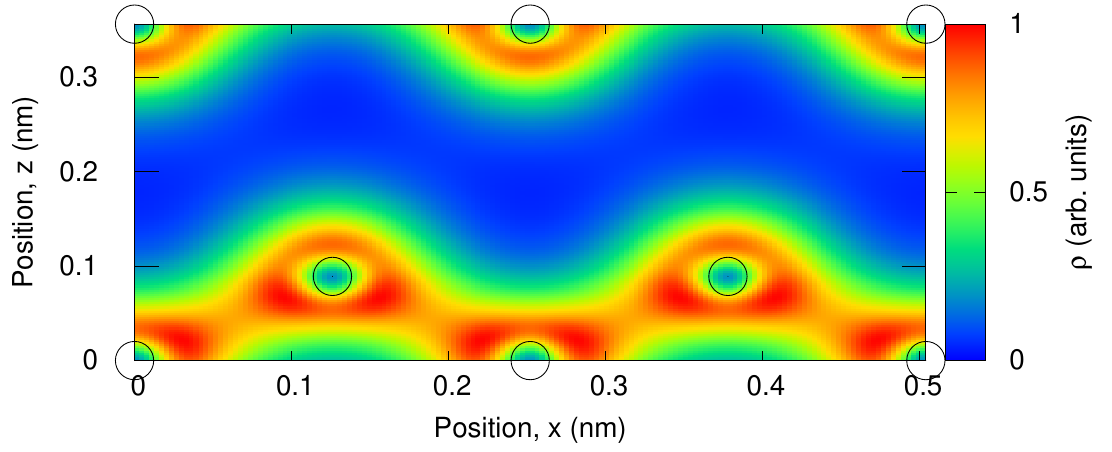}
\caption{\label{fig:rho_gs_diamond}
Ground state electron density $\rho(\vecb r)$ of diamond in the $\left  ( 110 \right )$ plane. The black circles indicate the positions of diamond atoms. Note that contributions only from valence electrons are included, but the core-electron contribution is excluded due to the pseudopotential approximation.
}
\end{figure}

To gain microscopic insight into the electronic properties of diamond, we revisit the electron density. Figure~\ref{fig:rho_gs_diamond} shows the ground state electron density in diamond. The covalent bonds are found to be formed between carbon atoms. Although both diamond and silicon show clear covalent bonds, electrons are more localized around the atoms than the bond regions in the case of diamond, whereas more significant electron localization can be found around the bond regions in the case of silicon. Hence, compared with the case of silicon, localized electron dynamics around carbon atoms are expected to be more important in diamond rather than the bond region.

\begin{figure}[htbp]
  \centering
  \includegraphics[width=0.95\columnwidth]{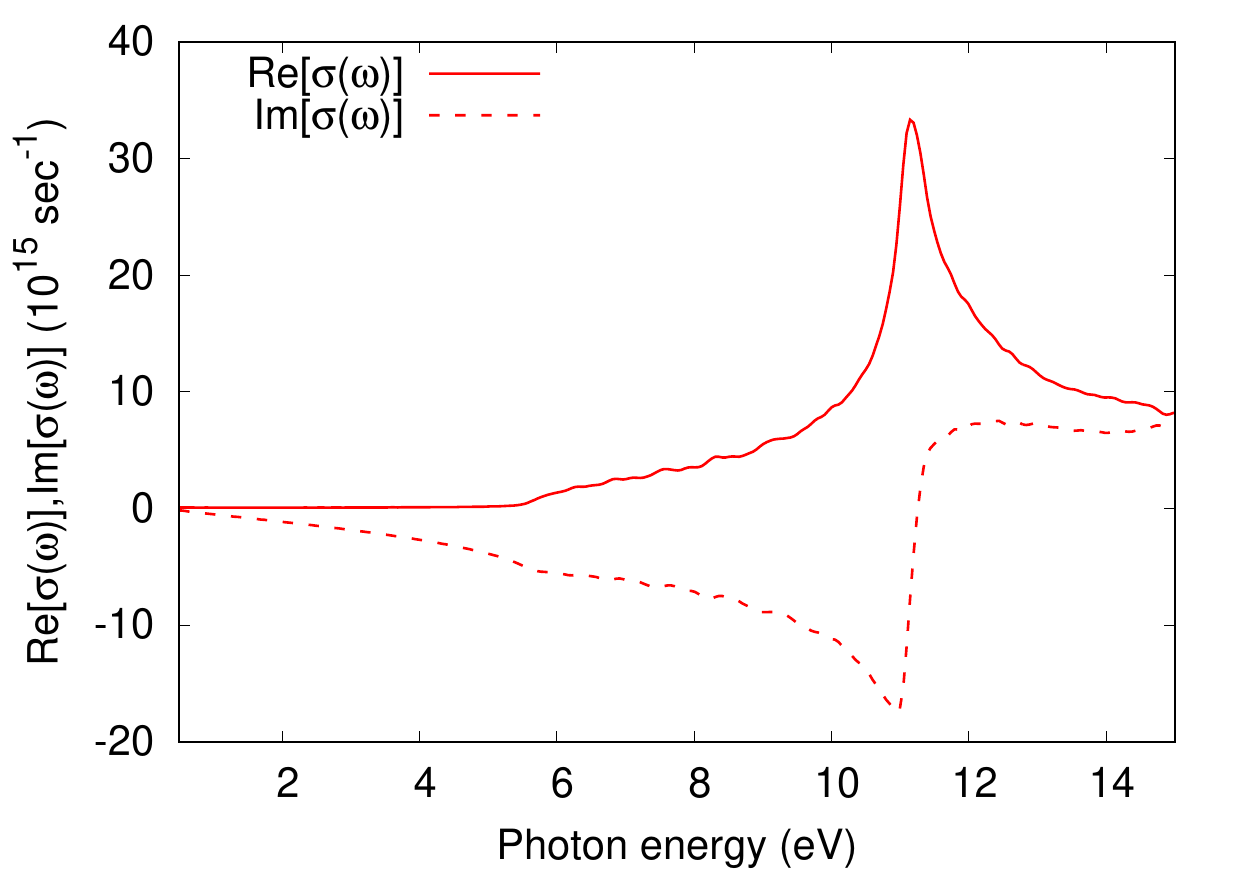}
\caption{\label{fig:sigma_diamond}
Optical conductivity, $\sigma(\omega)$, of diamond computed by TDDFT.
}
\end{figure}

Before investigating the nonlinear optical responses of diamond, we revisit the linear optical properties. Figure~\ref{fig:sigma_diamond} shows the optical conductivity of diamond computed via TDDFT. The computed optical gap of diamond is approximately $5.5$~eV, which is underestimated from the experimental gap of $7.1$~eV \cite{PhysRevB.46.4483}, and the real part of the conductivity is almost zero below the optical gap. In contrast to the real part, the imaginary part shows finite negative values below the optical gap. These behaviors are consistent with those of silicon in Fig.~\ref{fig:sigma_Si}.

As an example of the nonlinear optical effects, here we investigate the third harmonic generation from diamond. For this purpose, we compute the electron dynamics under a vector potential described by Eq.~(\ref{eq:vec-pot-al}) by setting $\omega_L$ to $1.0$~eV/$\hbar$, $E_0$ to $8.7\times 10^8$~V/m, and $T_{\mathrm{pulse}}$ to $20$~fs. Note that the corresponding three-photon energy is $3\omega_L \hbar=3.0$~eV, and it is below the optical gap of $5.5$~eV. Hence, the three-phonon process is still an off-resonant process under the present condition.

\begin{figure}[htbp]
  \centering
  \includegraphics[width=0.95\columnwidth]{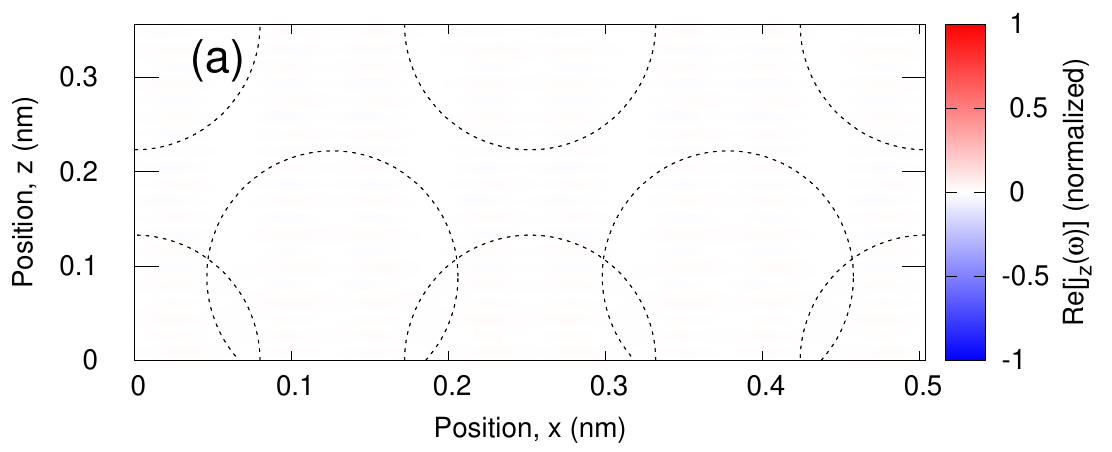}
  \includegraphics[width=0.95\columnwidth]{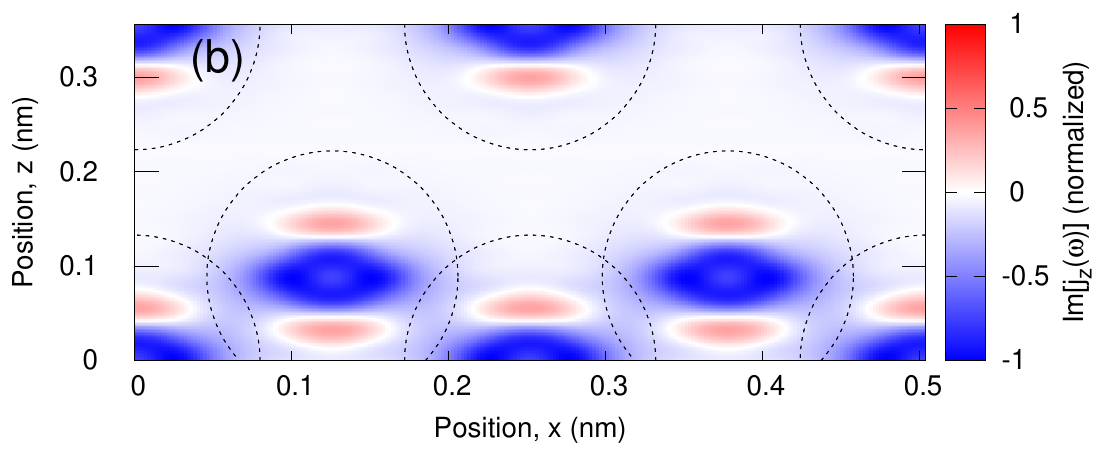}
\caption{\label{fig:zjz_re_w_diamond}
Frequency-resolved microscopic current density in diamond: $\tilde{j}_z(\vecb r,\omega)$. The real part is shown in (a), while the imaginary part is shown in (b). The corresponding electron dynamics is computed under the vector potential described by Eq.~(\ref{eq:vec-pot-al}). Here, the mean photon energy of the field $\hbar \omega_L$ is set to $1.0$~eV, and the frequency-resolved microscopic current density is evaluated at the frequency of $3\omega_L$ to elucidate the third-order harmonic generation. Here, $\tilde{j}_z(\vecb r,3\omega_L)$ is normalized by the maximum value of $\left |\tilde{j}_z(\vecb r,3\omega_L) \right |$ in the entire space. Here, the cutoff radius of the nonlocal pseudopotential is depicted as black dashed lines.
}
\end{figure}

Figure~\ref{fig:zjz_re_w_diamond} shows the $z$-component of the frequency-resolved microscopic current density, $\tilde{\vecb j}(\vecb r,\omega=3\omega_L)$, in diamond at the frequency of $3\omega_L$, which is the three times of the driving laser frequency. Hence, the microscopic current density in Fig.~\ref{fig:zjz_re_w_diamond} reflects the nature of the third-order harmonic generation. Similar to the cases of the linear responses of aluminum in Sec.~\ref{subsec:aluminum} and silicon in Sec.~\ref{subsec:silicon}, the real and imaginary parts of the frequency-resolved microscopic current density, $\tilde{\vecb j}(\vecb r,\omega)$, at $\omega=3\omega_L$ correspond to the real and imaginary parts of the nonlinear conductivity tensors, $\sigma^{(3)}_{ijkl}$, respectively. Reflecting the fact that the real part of the nonlinear conductivity is zero due to the off-resonant condition, the real part of the frequency-resolved microscopic current density in Fig.~\ref{fig:zjz_re_w_diamond} is almost zero in the entire space. In contrast, the imaginary part shows the localized structures around diamond atoms. This indicates that the third harmonic generation in diamond in the off-resonant regime is caused by the dynamics of bound electrons around carbon atoms rather than the covalent bond region. Furthermore, more microscopic insight is obtained: a positive contribution (red) appears around the relatively outer region of the bound electron dynamics, while a negative contribution (blue) appears around the inner region. Hence, there is an internal cancellation in the third-order harmonic generation in diamonds. Recently, a similar cancellation but in $k$-space has been discussed for the high-order harmonic generation in graphene \cite{PhysRevB.103.L041408}. Furthermore, the enhancement of the high-order harmonic generation has been proposed by suppressing the cancellation with an external degree of freedom. By extending the knowledge of the previous work, one may further enhance the third-order harmonic generation by suppressing the cancellation with an external degree of freedom, such as the secondary light field.

At the end of this section, we discuss a contribution from the nonlocal part of pseudopotentials to the macroscopic current density. The microscopic current density in Eq.~(\ref{eq:micro-current}) is defined only by the local part of the velocity operator in Eq.~(\ref{eq:velocity-local}), ignoring the nonlocal part in Eq.~(\ref{eq:velocity-nonlocal}). Hence, the spatial average of the microscopic current may not fully reproduce the macroscopic current density $\vecb J(t)$ if the nonlocal part of the velocity operator exists due to the pseudopotential approximation. Nevertheless, the contribution from the nonlocal part of pseudopotentials does not prevent the extraction of microscopic insight at the atomic scale. Because the nonlocal part of pseudopotentials is constructed with atomic orbitals and is localized within a cutoff radius, the contribution from the nonlocal part to the macroscopic current density is highly localized at atoms, indicating that the nonlocal contribution can be assigned to highly localized atomic responses.

Practically, the contribution from the nonlocal part of the velocity operator to the macroscopic current is often small. To assess the significance of the nonlocal contribution, we compute the power spectrum of emitted light from diamond by applying the Fourier transform to the macroscopic current density obtained by the above calculation as
\begin{align}
I(\omega) \sim \omega^2 \left |\int^{\infty}_{-\infty}dt W(t)e^{i\omega t}\vecb J(t) \right |^2.
\label{eq:power-spec}
\end{align}

\begin{figure}[htbp]
  \centering
  \includegraphics[width=0.95\columnwidth]{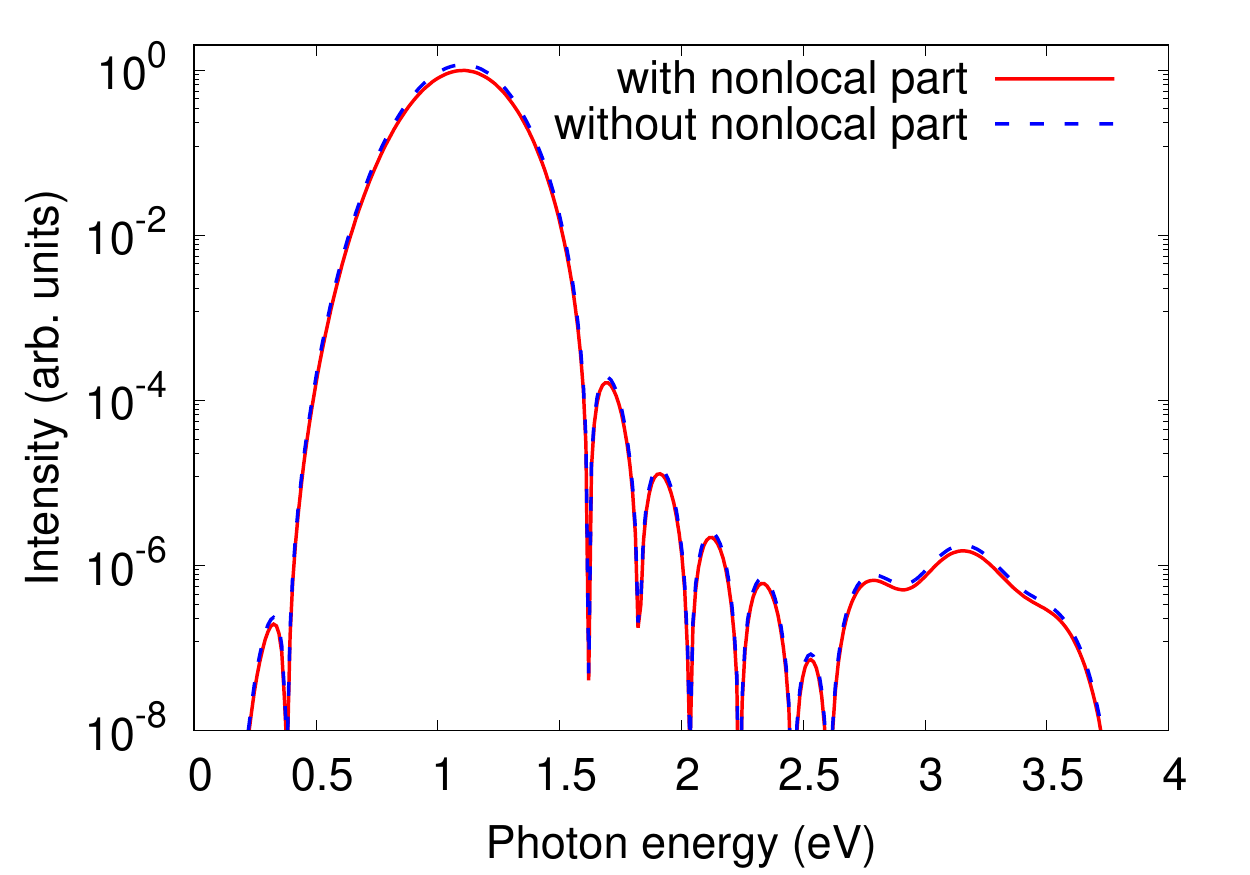}
\caption{\label{fig:harmonics_diamond}
Power spectrum, $I(\omega)$, obtained using the macroscopic current density in diamond with Eq.~(\ref{eq:power-spec}). The result computed by including the nonlocal velocity contribution is shown as the red solid line, whereas that computed by excluding the nonlocal contribution is shown as the blue dashed line.
}
\end{figure}

Figure~\ref{fig:harmonics_diamond} shows the computed power spectrum, $I(\omega)$. The red solid line shows the result obtained using the macroscopic current density with the nonlocal part of the velocity operator, whereas the blue dashed line shows the result without the nonlocal contribution. As seen from Fig.~\ref{fig:harmonics_diamond}, both responses at the fundamental frequency $\omega=1$~eV/$\hbar$ and the third harmonic frequency $\omega=3$~eV/$\hbar$ are well described, even without the contribution from the nonlocal velocity operator.

We evaluate the following quantity to quantify the error due to the lack of nonlocal contribution:
\begin{align}
I_{\mathrm{Error}}=\frac{\left |\int^{\infty}_{-\infty}dt W(t)e^{i\omega t}
\left \{\vecb J_{\mathrm{w/o~nonloc}}(t)-\vecb J_{\mathrm{w/~nonloc}}(t)\right \} \right |}
{
\left |\int^{\infty}_{-\infty}dt W(t)e^{i\omega t} \vecb J_{\mathrm{w/~nonloc}}(t) \right |
},
\end{align}
where $\vecb J_{\mathrm{w/~nonloc}}(t)$ is the macroscopic current density computed by including the contribution from the nonlocal part of the velocity operator, and $\vecb J_{\mathrm{w/o~nonloc}}(t)$ is the macroscopic current density computed by excluding the nonlocal contribution. The computed error $I_{\mathrm{Error}}$ is $8.1\times 10^{-2}$ at $\omega=3\omega_L$. Hence, in the present analysis of the third harmonic generation, we can safely ignore the nonlocal contribution in the velocity operator.

\section{Summary \label{sec:conclusion}}

In this study, we developed the Fourier analysis of the light-induced microscopic current density for investigating linear and nonlinear optical phenomena in solids. The frequency-resolved microscopic current density, $\tilde{\vecb j}(\vecb r,\omega)$, in Eq.~(\ref{eq:freq-current-density}) contains microscopic information regarding light-induced electron dynamics at a specific frequency of optical phenomena, providing real-space insight at the atomic-scale.

We first applied the developed approach to a linear response of metallic systems, considering aluminum as an example. As a result, we found that the frequency-resolved microscopic current density, in Fig.~\ref{fig:zjz_re_w_Al}, suitably captured the delocalized nature of the free electron dynamics in metals. We subsequently applied frequency domain analysis to a linear response of semiconductors in the off-resonant regime, considering silicon as an example. In contrast to metallic systems, the frequency-resolved microscopic current density clearly captures the bound electron dynamics around the covalent bonds in silicon, as shown in Fig.~\ref{fig:zjz_re_w_Si}. Therefore, the frequency-resolved microscopic current density analysis provides a comprehensive description for both delocalized free carrers in metals and bound electrons in semiconductors. Furthermore, we applied frequency domain analysis to nonlinear responses, considering the third harmonic generation from diamond as an example. We demonstrated that the third harmonic generation from diamond is caused by the bound electron dynamics around carbon atoms rather than the electron dynamics around covalent bonds.

The frequency-resolved microscopic current density analysis provides atomic-scale real-space insights into linear and nonlinear optical phenomena in solids. In addition to the traditional understanding of optical phenomena based on the $k$-space description, the real-space insights offer a complemental view of the phenomena. Hence, the Fourier analysis developed in this study may be used to strengthen the microscopic understanding of optical phenomena with more intuitive descriptions. For instance, one may identify responsible chemical elements or bonds for a specific optical response and use the knowledge to enhance it. Furthermore, the real-space description of the frequency-resolved microscopic current density does not rely on any basis expansion; rather, it naturally provides a real-space distribution. Hence, it may open a path for the investigation of highly nonlinear optical phenomena in nonperturbative regimes, where the $k$-space description based on the perturbation theory still poses fundamental difficulty.

\begin{acknowledgments}
This work was supported by JSPS KAKENHI Grant Numbers JP20K14382 and JP21H01842. The author thanks the Supercomputer Center, the Institute for Solid State Physics, the University of Tokyo for the use of the facilities.
\end{acknowledgments}

\bibliography{ref}

\end{document}